# The putative nebula of the Wolf-Rayet WR 60 star: A case of mistaken identity and reclassification as a new supernova remnant G310.5+0.8

M. Stupar[1,2] • Q.A. Parker[1,2] • M.D. Filipović [3]

[1]Department of Physics, Macquarie University, Sydney 2109, Australia
[2]Anglo-Australian Observatory, P.O. Box 296, Epping, NSW 1710, Australia
[3]University of Western Sydney, Locked Bag 1797, Penrith South DC, NSW 1797, Australia

**Abstarct**

We present narrow band AAO/UKST H$\alpha$ images and medium and low resolution optical spectra of a nebula shell putatively associated with the Wolf-Rayet star WR 60. We also present the first identification of this shell in the radio regime at 843 MHz and at 4850 MHz from the Sydney University Molonglo Sky Survey (SUMSS), and from the Parkes-MIT-NRAO (PMN) survey respectively. This radio emission closely follows the optical emission. The optical spectra from the shell exhibits the typical shock excitation signatures sometimes seen in Wolf-Rayet stellar ejecta but also common to supernova remnants. A key finding however, is that the WR 60 star, is not, in fact, anywhere near the geometrical centre of the putative arcuate nebula ejecta as had been previously stated. This was due to an erroneous positional identification for the star in the literature which we now correct. This new identification calls into serious question any association of the nebula with WR 60 as such nebula are usually quite well centred on the WR stars themselves. We now propose that this fact combined with our new optical spectra, deeper H$\alpha$ imaging and newly identified radio structures actually imply that the WR 60 nebula should be reclassified as an unassociated new supernova remnant which we designate G310.5+0.8.

## 1 Introduction

The morphological formation and optical spectral characteristics of different emission structures can be divided into several main groups: remnants of supernova explosions, planetary nebulae (PN), H II regions and Wolf-Rayet shells. Although the roots of these nebulosities are in very different physical processes and are manifestations of objects of very different mass and evolutionary state they can often exhibit strong morphological similarities in the form of nebula shells or semi-circular or filamentary structures as well as showing strong similarities in their optical spectra.

In this paper we are concerned with the nebula shell previously identified with the Wolf-Rayet star WR 60. These usually extensive filamentary emission nebulae are the product of massive Wolf-Rayet stars. Such

[1,2,3] mstupar@physics.mq.edu.au ; qap@physics.mq.edu.au ; m.filipovic@uws.edu.au

~ 40 $M_\odot$ evolved, extremely hot and luminous stars undergo rapid mass loss of the order of $5\times10^{-5}$ $M_\odot$ yr$^{-1}$ forming both, shock-excited and photoionised nebula around them (Johnson and Hogg). A Wolf-Rayet star should always be found close to or at the centre of such ejected shells. These shells are not always spherical in nature and there is variety of morphological elliptical and asymmetrical ring forms due both to the evolutionary stage of the ejecta and the interactions the ejecta may have had with the surrounding ambient ISM [and references therein]grue00. There is also some confusion between the optical spectral signatures of Wolf-Rayet shells and supernova remnants (SNRs). This is especially in the observed ratios of the [S II] to H$\alpha$ lines, where typical values of  [S II] / H$\alpha$ > 0.5, indicative of shock excitation, are found for SNRs (enabling discrimination from from H ll regions and most of PN) but is also found in some Wolf-Rayet shells (see example in Esteban et al. 1990). Indeed, although a variety of spectral line diagnostics are currently used to quite effectively separate most SNRs, PN, Wolf-Rayet shells and H ll regions, extreme examples of each class overlap in the various emission line ratio diagnostic diagrams as shown by Sabbadin, Minello and Bianchini (1977), Cantó (1981) and most recently and clearly by Frew and Parker (2010). However, we can also perform radio observations of Wolf-Rayet shells to provide extra diagnostic capability as, unlike SNRs, Wolf-Rayet shells are typically thermal radio sources, with $\alpha \approx$ -0.1 ( $S_\nu \propto \nu^\alpha$ ). For SNRs $\alpha \geq -0.5$ is expected as they are strong non-thermal sources with synchrotron emission (Filipović et al.1998). However, some remnants also exhibit $\alpha$ values similar to those found for Wolf-Rayet shells (see review of fluxes in Green 2009). Unfortunately, this can sometimes also make classification of SNRs and Wolf-Rayet shells via radio criteria difficult in the absence of other indicative evidence. Another possible distinction between Wolf-Rayet shells and possible contaminants is detection of H I 21 cm line radio emission in the extensive bubbles present around massive stars (as found currently around 35 Wolf-Rayet stars; see review in Cappa 2006. These bubbles consist of neutral gas further out from the ejected optical emission nebula shell (Cappa 2006).

Of course any ambiguity concerning the nature of the emission structure is immediately solved if the progenitor Wolf-Rayet star is identified close to the geometric centre of the nebula. This is what has long been considered the case regarding the nebula identified with the WC7 Wolf-Rayet star WR 60 until now. We have noticed a positional irregularity in previous work that calls into question the assigning of this star to the nebula (see below).

For the first time we also show newly obtained spectra for the nebula component previously identified with the WR 60 star, together with high resolution, high sensitivity narrow-band images of this nebula taken from the SuperCOSMOS H$\alpha$ survey of the southern Galactic plane (SHS) (Parker et al. 2005). We also show radio images of the region uncovered in the SUMSS 843 MHz sky survey (Bock, Large and Sadler 1999) and PMN 4850 MHz survey (Condon, Griffith and Wright 1993) which have a strong positional coincidence with the H$\alpha$ observations. We analyse our spectra and images and discuss options regarding the likely SNR nature of the nebula previously associated with the WR 60 star and finally suggest some future work.

## 2  Observing results for the nebula in the vicinity of the WR 60 Star

Marston et al. (1994) presented the original discovery of a putative nebula around the Wolf-Rayet star WR 60. This was based on a 900 second narrow-band H$\alpha$ CTIO 0.6m Curtis-Schmidt CCD image. However, Marston et al. (1994) had incorrectly identified the Wolf-Rayet star on their Fig. 7. The marked star in their figure is actually a much brighter (V=9.83) star TYC 9004-2733-1 (GSC 09004-02733) at RA(2000)=13$^h$55$^m$43.2$^s$ and δ=-61°04'14". The true position of the Wolf-Rayet WR 60 star (also known as HD121194; V=12.17 - taken from SIMBAD), is actually RA(2000)=13$^h$55$^m$48.1$^s$ and δ=-61°09'50" approximately 4 arcminutes to the south (as given in the van der Hucht 2001). This is clearly shown in Fig. 1.

It is now quite evident that the correctly identified WR 60 star is not situated anywhere near the geometric centre (radius of curvature) of the surrounding arcuate H$\alpha$ nebula (see our Fig. 1 where both stars are clearly marked) but is offset by approximately 4 arcminutes to the south-east. A reasonable central location for the Wolf-Rayet star within the nebula is expected and is the case among other known WR stars with associated nebula (see Chu and Lasker 1980). The position of the now correctly identified WR 60 star is also shown on Fig. 3 in relation to radio image of this shell.

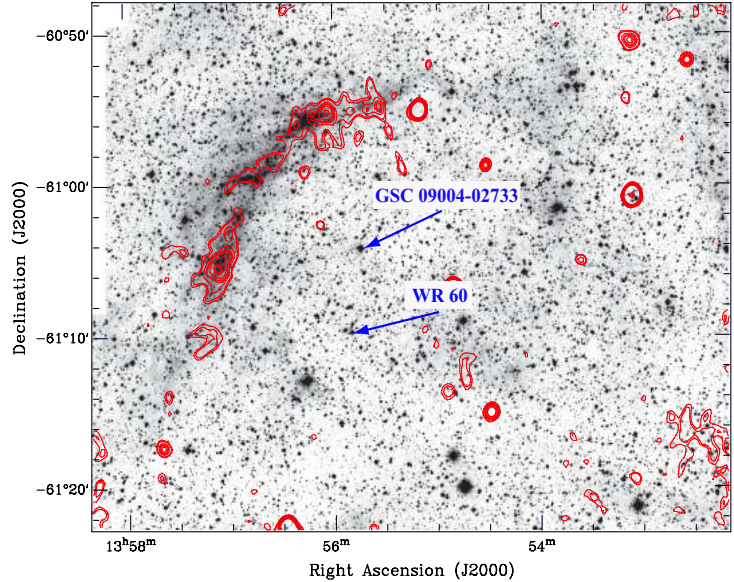

**Figure 1**: The arcuate nebulosity putatively associated with the WR 60 star in Hα light (mosaic image) overlaid with SUMSS radio contours (from 0.006 to 0.01 Jy Beam$^{-1}$). There is an excellent match between the optical and radio images. The arrows also give the true position of the Wolf-Rayet star WR 60 (HD 121194) as RA(2000)=13$^h$ 55$^m$ 48.1$^s$; δ = -61° 09' 50" as well as the position of the star GSC 09004-02733 previously identified in Marston et al. (1994) as the WR 60 star. From this it is clearly seen that WR 60 is not in fact close to the centre of the optical and radio arcuate structures.

Additionally, Marston et al. (1994) comment that this nebula has a ring that extends almost completely (90%) around this star as might be expected for a true Wolf-Rayet shell nebula. This ring is not evident in their Fig. 7 which only shows a nebula arc about 20 arcminutes in extent to the NE of the incorrectly identified WR 60 star. Furthermore, there is no such complete ring obvious from our new, deeper, higher quality Hα images though the strong arcuate structure in the NE is confirmed. However, more diffuse and much fainter irregular shaped nebulosity can be seen extending to the NW and SW in our quotient imaging, particularly a clump to the SW. Quotient imaging is better at revealing lower surface brightness features (left panel on Fig. 2) where a fragmented low level ring structure could in fact be present but at much lower intensity and coherence than the strong NE arc. Note the SHS Hα survey (Parker et al. 2005) has arcsecond resolution and 5 Rayleigh sensitivity and is deeper than the Marston et al. (1994) data.

The main NE arc consists of several apparently overlapping broad emission filaments (Figs. 1 and 2) with an excellent match to radio emission in the SUMSS 843 MHz image (right panel Fig. 2). The SUMSS image clearly displays a strong arcuate radio structure co-incident with the optical nebulosity and which is very similar in size (~20) and morphological structure to that seen in Hα. There is a further low level radio emission clump to the SW that also matches the structure seen in the optical quotient image which might form part of a possible fragmented ring structure. In addition to the SUMSS data, the PMN radio survey data at 4850 MHz also detected radio emission co-located with the SUMMS data and optical emission but, due to the lower resolution of this survey, is seen in the form of 3 fragmented and broader areas of radio emission that do seem to join-up form an overall ring structure but with an opening to the NW. The peak flux is 0.18 Jy Beam$^{-1}$ (see Fig. 3).

If there are doubts concerning whether the SW emission clump seen in the SUMSS image is part of a possible radio ring structure or is completely separate, we can see that the PMN 4850 MHz flux in this area is stronger than for the weak NE radio arc (peak 0.20 Jy Beam$^{-1}$). Besides, while this NE structure is the same size as seen in the optical (Hα or quotient Hα and SR image) and SUMSS 843 MHz radio map (~6'), its PMN size is ~13.5×8.5'.

Finally, the available (low resolution) IRAS and MSX mid-infrared data does not show any obvious ring structure. Though dust structures are present they do no appear to be not connected with our optical and radio structures nor do they outline any shell.

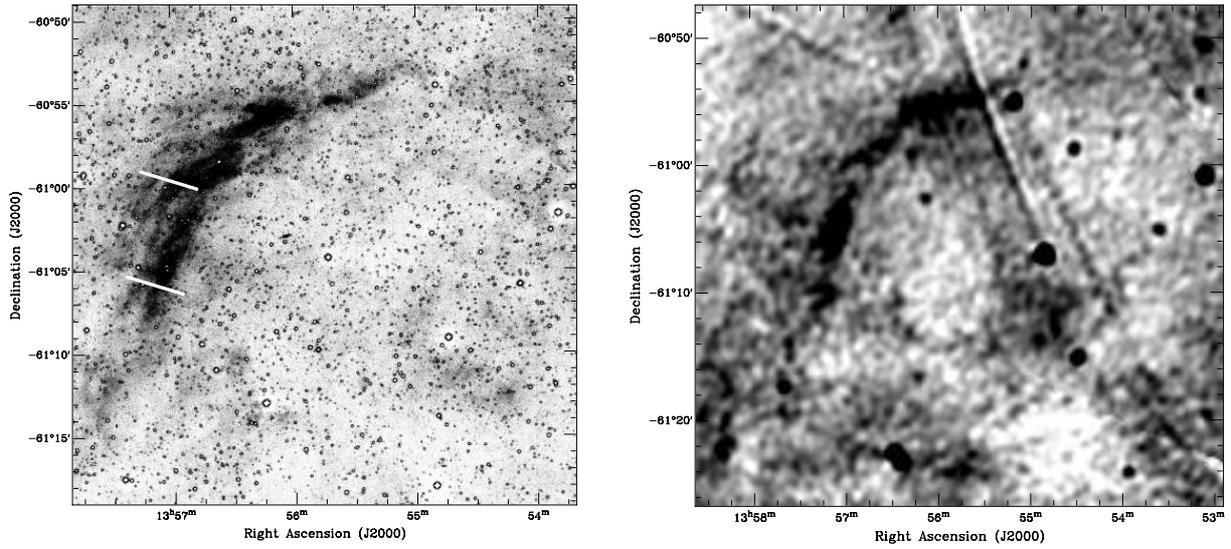

**Fig. 2:** The left figure is the SHS quotient obtained by dividing the Hα and short red (SR) images of the nebula previously identified as emanating from the WR 60 star which shows up the emission more clearly than the direct Hα image in Fig. 1. The slit positions for our spectral observations are also indicated. The right hand figure is the same field from the SUMSS 843 MHz radio survey which clearly displays a strong arcuate radio structure co-incident with the optical nebulosity and which is very similar in angular extent (~20') and morphological structure to that seen in Hα. However, additional low level emission is seen on both images (slightly brighter in the radio) to the SW that might be associated and could form part of a possible overall ring structure. See also Fig. 3.

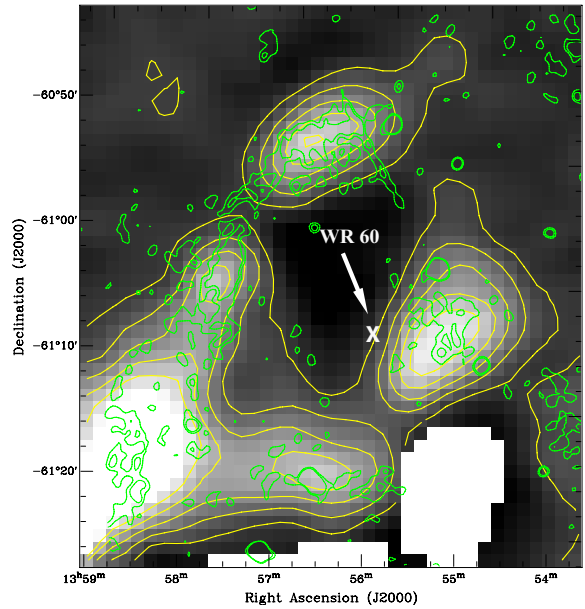

**Figure 3:** The Parkes-MIT-NRAO (PMN) image of the putative WR 60 shell at 4850 MHz. The yellow colour contours show flux values at this frequency, from 0.01 to 0.2 Jy Beam$^{-1}$ overlaid with SUMSS 0.004 to 0.01 Jy Beam$^{-1}$ green colour contours of the same area shown on the right panel of Fig. 2. Due to the higher resolution of the SUMSS survey (~43 arcsec) the radio structure at this frequency is more fragmented compared with the PMN radio structure. Nevertheless, two strong localised radio emissions in the NE region can be noticed in the PMN map (peak flux 0.18 Jy Beam$^{-1}$) with an excellent match to the SUMSS 843 MHz contours (from Fig. 1). An excellent match with the SUMSS image can be also noticed to the SW with the peak value at 4850 MHz of 0.20 Jy Beam$^{-1}$, which at this frequency is stronger than the prominent NE arc. The **X** marks the corrected position of the WR 60 star.

# 3 Method and Observations

Spectra of the nebula previously identified as ejecta from the Wolf-Rayet star WR 60 were obtained with two telescopes and spectrographs in June and July 2004 (see Table 1). The first was the Double Beam Spectrograph (DBS)[2] of the Mount Stromlo and Siding Spring Observatory 2.3m telescope. The DBS had a dichroic which splits the visible light into components which feed the blue and red arms of the spectrograph. For the blue we used a 600 lines mm$^{-1}$ grating which covered the region from 3700 and 5500Å. For the red arm we used a higher resolution 1200 lines mm$^{-1}$ grating covering the region between 6100 and 6800Å which samples the important diagnostic nebula lines in the red and is also of sufficient resolution for meaningful kinematics to be determined. The slit width was 2.5" and a resolution of 2 and 1Å was achieved in the blue and red arms respectively. The second spectrograph[3] was attached to the 1.9-m Radcliffe telescope of the South African Astronomical Observatory (SAAO). Here we used a 300 lines mm$^{-1}$ grating with a resolution of ~7Å, ideal for the initial spectral classification of the nebula spectra as it covered a broad spectral range between 3500 and 7500Å. Observational details are given in the Table 1.

Data reduction was performed using standard *IRAF* spectral reduction routines. For spectral flux calibration we used observations of the standard photometric stars: LTT4302 for June 14, 2004 and LTT6248 for the July 22, 2004 observations.

**Table 1**: Spectral observation log for the various observed optical components of the putative WR 60 star nebulae. Each exposure was for 1200 seconds.

| Object (nebula) | Telescope | Date | Grating (lines mm$^{-1}$) | Spectral range (Å) | Slit RA h m s | Slit Dec. ° ′ ″ |
|---|---|---|---|---|---|---|
| WR 60 | 2.3-m | 14/06/2004 | 600$^a$ | 3700–5500 | 13 57 05 | –61 05 20 |
|  | 2.3-m | 14/06/2004 | 1200 | 6100–6800 | 13 57 05 | –61 05 20 |
|  | 1.9-m | 22/07/2004 | 300$^b$ | 3500–7500 | 13 56 49 | –60 59 31 |

$^a$For the 600 and 1200 lines mm$^{-1}$ gratings the rms error in the dispersion solution (in Å) was between 0.05 and 0.02 and the relative percentage error in the flux estimate was between 10% and 17% for the 600 lines mm$^{-1}$ grating and ~20% for the 1200 lines mm$^{-1}$ grating.

$^b$For the 300 lines mm$^{-1}$ grating on the SAAO 1.9m the rms dispersion error (in Å) was 0.03 and the relative percentage error in the flux estimate was ~6%.

## 3.1 Optical spectra of the nebula

For the first time we show optical spectra taken at two locations across the nebula clearly shown at Fig. 2. Let us first examine the spectrum taken with the SAAO 1.9-m telescope (Fig. 4; see also Table 2) from July 22, 2004, as this one represents a complete optical spectrum from blue to red and is thus better for overall classification once flux-calibrated. The general characteristics of SNR optical emission lines are exhibited, with the ratio of [S II] / Hα =0.58 which eliminates any possible confusion with a H ll region. Such a high ratio is also extremely rare in PN but does fall (just) within extreme values found for some highly evolved PNe interacting with the ISM (Pierce et al. 2004; Frew and Parker 2010). However, in the blue, Hβ and Hδ as well as [O II] at 3727Å can be seen but with no detectable [O III] at 5007Å. The lack of [O III] at 5007Å now effectively rules out a PN origin for the nebula. Unfortunately, the observed ratio of the weak [O II] 6300 and 6364Å lines is uncertain due to the non-photometric night and difficulty with proper sky-subtraction in the lower resolution, low S/N spectrum, although their detection also supports an SNR origin.

A few weeks prior to the SAAO observation, we observed across the main NE arcuate nebula on June 14, 2004 with the MSSSO 2.3-m DBS spectrograph, some 5′ further south from the SAAO position (see Fig 2) where the blue and red spectral regions are covered separately at higher resolution than for the SAAO data. The reduced spectrum gave [S II] / Hα =0.78, somewhat higher then the SAAO value and even more firmly in the domain typical of SNRs but not H II regions or PN (see Fig. 6 and Table 2). Here, the [O I] lines at 6300 and 6364Å are more clearly present, again supporting SNR classification. Unfortunately the lines are again not in their expected atomic ratio 3:1 due to difficulties in proper sky subtraction where these lines are very strong in comparison. In the blue DBS spectrum (Fig. 5) the Balmer lines can be seen, with strong Hβ, and clearly seen at 3727Å. Both spectra fit a SNR classification. As the blue DBS spectrum and that from the SAAO 1.9-m have low S/N, better spectra are required. However, the spectral evidence for an SNR identification is already compelling as the observed detected line ratios are already sufficient to eliminate confusion with a H ll region or PN. This leaves only the possibilities of an SNR or WR ejecta as the most likely origin for the observed nebula.

---

[2] Now obsolete. See http://msowww.anu.edu.au/observing/ssowiki/index.php/2.3m_DBS
[3] http://www.saao.ac.za/facilities/instrumentation/gratingspec/

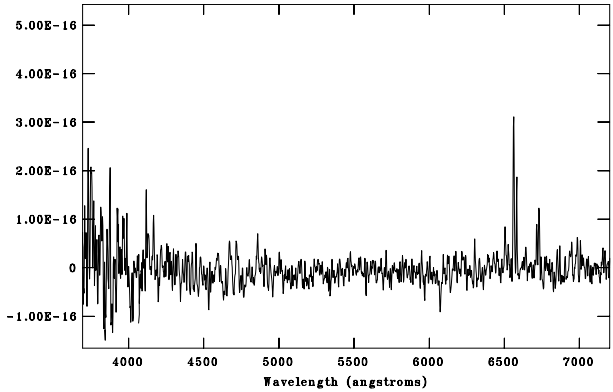 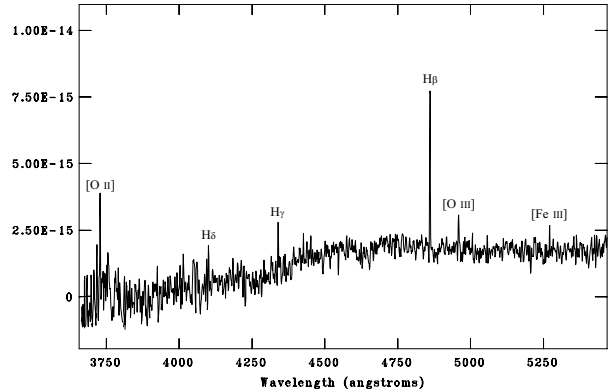

**Figure 4:** The flux calibrated though low S/N SAAO 1.9m spectrum of the putative nebula of the WR 60 star in the range between 3500 and 7500Å taken with a low resolution 300 l mm$^{-1}$ grating. This spectrum is from the slit position at RA(2000)=13$^h$56$^m$49$^s$ and δ=-60°59'31" from July 22, 2004. The low dispersion and non-photometric night produced a noisy spectrum. Nevertheless, the most important lines for classification can be seen, from the strong [S II] 6717 and 6731Å emission (which rules out a H II region) the [N II] at 6548 and 6584Å emission lines in the red to the Balmer lines and [O II] at 3727Å in the blue. Note the absence of [O III] at 5700 Å which also rules out a PN origin for the nebula.

**Figure 5:** The flux calibrated blue spectrum of the nebula obtained with the MSSSO 2.3-m DBS spectrograph at RA(2000)=13$^h$57$^m$54$^s$ and δ =-61°05'19". Although the spectrum between 3700 and 4500Å has low S/N, the [O II] emission line at 3727Å can be noticed as well as all Balmer lines and the absence of detectable [O III] at 5007Å. The possible weak detection of [Fe III] at 5270Å with a ratio of 14 against Hβ is a strong supporter of a SNR origin of this spectrum.

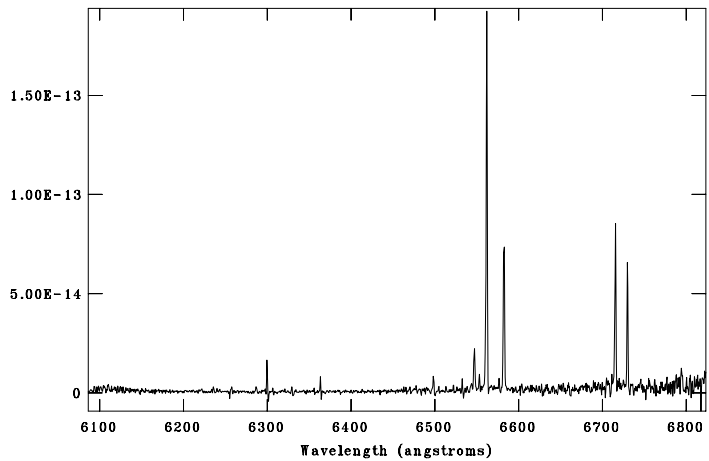

**Figure 6:** The better quality, higher dispersion, flux calibrated red spectrum obtained for the same slit position as for the Fig. 5 from the MSSSO 2.3-m DBS. The ratio of [S II] / Hα =0.68, is typical for an SNR. Oxygen lines of [O I] at 6300 and 6364Å can be noticed in the expected 3:1 ratio and are also typically seen in SNRs.

**Table 2:** Optical emission lines detected in the red and blue MSSSO 2.3-m DBS spectra from June 14, 2004 and SAAO 1.9m telescope low resolution (300 lines mm$^{-1}$ grating) observations from July 22, 2004.

| Line | λ (Å) | F(λ) Hβ=100 | Line | λ (Å) | F(λ) Hα=100 |
|---|---|---|---|---|---|
| **MSSSO observations** | | | | | |
| [O II] | 3727 | 63 | [O I] | 6300 | 9 |
| Hδ | 4101 | 24 | [O I] | 6364 | 4 |
| Hγ | 4340 | 32 | [N II] | 6548 | 11 |
| Hβ | 4861 | 100 | Hα | 6563 | 100 |
| [O III] | 4959 | 20 | [N II] | 6583 | 38 |
| [O III] | 5007 | | [S II] | 6717 | 44 |
| [Fe III] | 5271 | 14 | [S II] | 6731 | 34 |
| F(Hβ)=1.07×10$^{-14}$ erg cm$^{-2}$ s$^{-1}$ Å$^{-1}$ | | | F(Hα)=3.18×10$^{-13}$ erg cm$^{-2}$ s$^{-1}$ Å$^{-1}$ | | |
| [N II]/Hα=0.49 | | | | | |
| [S II]/Hα=0.78 | | | | | |
| [S II](6717/6731)=1.30 | | | | | |
| **SAAO observations** | | | | | |
| [O II] | 3727 | 665 | | | |
| Hδ | 4101 | 88 | | | |
| Hβ | 4861 | 100 | | | |
| [O I] | 6300 | 138 | | | |
| [N II] | 6548 | 49 | | | |
| Hα | 6563 | 747 | | | |
| [N II] | 6583 | 412 | | | |
| [S II] | 6717 | 147 | | | |
| [S II] | 6731 | 289 | | | |
| F(Hβ)=3.88×10$^{-16}$ erg cm$^{-2}$ s$^{-1}$ Å$^{-1}$ | | | | | |
| [N II]/Hα=0.62 | | | | | |
| [S II]/Hα=0.58 | | | | | |
| [S II](6717/6731)=0.51 | | | | | |

# 4 Discussion and conclusion

In the literature there is currently a putative shell nebula associated with the Wolf-Rayet star WR 60 e.g. Marston et al. (1994). Our re-evaluation of the positional relationship between WR 60 and nebula and our new multi-wavelength imaging and optical spectra now suggest this this association is erroneous. We summarise the evidence for this as it currently stands.

## 4.1 Positional arguments

Firstly a Wolf-Rayet star is expected to eject an essentially spherical shell or symmetric nebulae (Chu and Lasker 1980) centred on the star and indeed many known WR ejecta exhibit a complete or near-complete shell form with structures/filaments radiating out from the central star. In support of this Marston et al. (1994) reported a circular nebula shape on their CCD images apparently centred around WR 60. We find that the star identified as WR 60 at the centre of the putative shell by Marston et al. (1994) is actually a completely different star. We now confirm that the true Wolf-Rayet star WR 60 (HD 121194) is not situated anywhere near the geometric centre of the prominent, surrounding 20 arcminute arcuate nebula but is offset some 4 arcminutes to the south-east. This one key fact alone seriously undermines any association of the nebula and WR 60.

## 4.2 Morphological arguments

Classification of WR nebula (Chu 1981) via optical emission and morphology is connected with [O III] and [N II] imaging as these data provide global spectral information and they are good tracers of shock and ionization fronts (Chu 1981) as well as imaging in Hα light (which usually includes [N II] in the filter transmission). Our new, superior SHS Hα imaging does not show a complete nebula emission ring as reported by Marston et al. (1994) further undermining its identification as a bona-fide Wolf-Rayet shell though very low level optical emission is seen to the NW and SW. Hence, it is difficult to determine a position angle if the observed fraction of this putative ring is actually a projected ellipse with a sufficient opening angle to the line of sight to explain the existing offset position of the true WR 60 star from the centre of the nebula. In the PMN radio observations (Fig. 3), a near complete ring is apparent. The optical filamentary structure of the nebula is also typical of SNRs which is further corroborated by the positional match to the radio data which, as it peeks through the dust, is also able to reveal more of the true ring structure typical of many SNRs.

### 4.3 Spectral arguments

Inspection of our spectra shown in Figs. 4, 5 and 6 and the emission line ratios listed in Table 2 shows that [O III] emission was not seen in our spectra (see excellent example of [O III] dignostic of Wolf-Rayet shells in Grundel et al. 2000). This does not support identification as a true Wolf-Rayet emission shell where such emission is usually strong. The observed optical emission line ratios also rule out any possible confusion with a PN or H II region. However, the observed high [S II] to H$\alpha$ ratio is indicative of shock excitation, typical of SNRs, while the clear detection of [O I] 6300 and 6364Å in the higher S/N DBS red data also supports an SNR identification.

### 4.4 Other arguments

Some contradiction exists with Nichols and Fesen (1994) where, in one possible SNR scenario, the progenitor star was part of a binary system with a WR star. In this situation a compact star should exist shortly after the supernova explosion with the WR star therefore also near the centre of SNR shell. This is again opposite to what is observed in the case of WR 60.

Additional radio observations at multiple frequencies could help in making a more definitive identification of this nebula as resulting from an SNR or WR star by revealing a thermal or non-thermal signature (WR nebula have thermal radio emission, e.g. Cappa 2006).

### Acknowledgments

We are grateful to the South African Astronomical Observatory and Mount Stromlo and Siding Spring Observatory Time Allocation Committees for enabling the spectroscopic follow-up to be obtained. We thank the WFAU of the Royal Observatory Edinburgh for the provision of the SHS data on-line.